\documentclass{xeus}
\usepackage{graphics}
\usepackage{psfig}

\begin{document}
\onecolumn
\title{A fast X-ray timing capability on XEUS}

\author{D. Barret\inst{1} \and G. K. Skinner\inst{1} \and
E. Kendziorra\inst{2} \and R. Staubert\inst{2} \and P. Lechner\inst{3} \and L. Str\"uder\inst{3}
\and M. van der Klis\inst{4}
\and L. Stella\inst{5}\and C. Miller\inst{6}}

\institute{ Centre d'Etude Spatiale des Rayonnements,
  CESR-CNRS/UPS,France (Didier.Barret@cesr.fr) \and University of Tubingen, Germany  \and MPI Halbleiterlator, Germany \and University of Amsterdam, The
  Netherlands \and University of Roma, Italy \and University of Maryland, United States}

\maketitle

\begin{abstract}
Fast X-ray timing can be used to probe strong gravity fields around collapsed objects and constrain the equation of state of dense matter in neutron stars. These studies require extremely good photon statistics. In view of the huge collecting area of its mirrors, XEUS could make a unique contribution to this field. For this reason, we propose to include a fast X-ray timing capability in the focal plane of the XEUS mirrors. We briefly outline the scientific motivation for such a capability. We compute some sensitivity estimates, which indicate that XEUS could provide better than an order of magnitude sensitivity improvement over the Rossi X-ray Timing Explorer. Finally, we present a possible detector implementation, which could be an array of small size silicon drift detectors operated out of focus. 
\end{abstract}

\section{Introduction}

The X-rays generated in the inner accretion flows around black holes
(BHs) and neutron stars (NSs) carry information about regions of the
strongly curved space-time in the vicinity of these objects. This is a
regime in which there are important predictions of general relativity
still to be tested. High resolution X-ray spectroscopy and fast timing
studies can both be used to diagnose the orbital motion of the
accreting matter in the immediate vicinity of the collapsed star,
where the effects of strong gravity become important. With the
discovery of millisecond aperiodic X-ray time variability (QPO) from
accreting BHs and NSs, and brightness burst oscillations in NSs, the Rossi X-ray Timing Explorer (RXTE, Bradt et al. 1993) has clearly demonstrated that fast X-ray timing has the potential to
measure accurately the motion of matter in strong gravity fields and
to constrain masses and radii of NSs, and hence the equation of state
of dense matter. 

Although the prime objectives of XEUS are to perform spectroscopy of faint X-ray sources to trace the origin and evolution of hot matter back to the early ages of the Universe (see Hasinger et al. 2000; Bleeker et al. 2000), the large collecting area required to meet these objectives could also be used for X-ray timing studies. In this paper, we briefly outline how XEUS could broaden its science by having a fast X-ray timing capability in the focal plane (section 2). The potential of X-ray timing studies is described in more details in two companion papers by Michiel van der Klis and Cole Miller. In section 3, we give some sensitivity estimates for XEUS. Finally, in section 4, we present a possible implementation for the fast X-ray timing capability.

\section{Summary of scientific objectives}
\label{summary_science}
\subsection{Probing strong gravity fields}
RXTE has detected high-frequency QPOs from both BH and NS binary systems (see van der Klis 2000 for an extensive review). In neutron stars, several types of QPOs are commonly observed,
including the two ``kilohertz" QPOs at $\nu_1=200-800$~Hz and
$\nu_2=500-1300$~Hz.  The upper end of this frequency range
represents the highest-frequency oscillations ever seen in
astronomy.  Indeed, the frequencies are so high that any
plausible mechanism requires that their origin be deep in
the gravitational well of the neutron star.  For example, the
orbital frequency 20~km from a $1.4\,M_\odot$ neutron star is
770~Hz, so the majority of the upper kilohertz QPOs must be
generated in regions of strong spacetime curvature.  As a
consequence, detailed study of these QPOs has great promise
for discoveries about strong gravity and the dense matter of
the neutron stars themselves.

The precise interpretation of the QPOs depends on the model
proposed.  Ideas include beat-frequency models (Miller, Lamb,
\& Psaltis 1998), where an orbital frequency is modulated by
emission at the stellar spin frequency; pictures in which the 
observed frequencies are related to geodesic frequencies modified by 
additional, possibly fluid, interactions (Stella \& Vietri 1998;
Psaltis \& Norman 2002); and disk oscillations (Titarchuk \&
Osherovich 1999).  Independent of the detailed origin of the QPOs,
there is wide agreement that the upper kilohertz frequency $\nu_2$
is close to the frequency of a nearly circular orbit at some special 
radius near the star.  From this alone one expects a number of
crucial phenomena to be detected with a high-area timing detector
on XEUS.  For example, an upper limit to $\nu_2$ is set by the
orbital frequency at the innermost stable circular orbit (ISCO).
With a highly sensitive detector, one expects to see a clear
upper limit to the frequency corresponding to the ISCO in several
sources.  Suggestive
evidence for this has been found for the source 4U~1820--30
(Zhang et al.~1998), but it is not yet conclusive.  Detection of
such a ceiling on the frequency will establish the existence of
unstable circular orbits, a fundamental prediction of general
relativity that is essential to the description of accreting black
holes of all masses as well as to neutron stars.  It will also allow
precise estimation of the mass of the neutron star, and is likely to
provide evidence for frame-dragging.  The detection of even higher 
frequencies than have yet been seen would allow the elimination
of several candidate equations of state of dense matter (Miller 
et al.~1998).  In addition, the detailed predictions of specific
models will be tested severely by the new data (e.g., Miller et al.
1998; Stella \& Vietri 1999).

Models for black hole QPOs will also be illuminated greatly by
high-area timing observations.  The detection of a pair of high-frequency
QPOs in three black hole candidates (Strohmayer 2001a,b; Miller et al.
2001), combined with their dynamical mass estimates, has already
provided clinching evidence for spinning black holes in these sources.
However, there is much ambiguity about the physical origin of these
oscillations.  A timing instrument with the area of XEUS would make
profound contributions to such study, not least because the oscillations
would be detectable on their coherence time, or even during a single
oscillation.  This will allow detailed characterization of the
brightness variations.  Specifically, because the cycle waveform
depends on the Doppler shifts associated with the local velocity of
the radiating matter in the emitting blob or spot, as well as on the
curved-spacetime light propagation effects, fitting of the waveform
yields the mass and spin of the compact object.  In fact, along with
detailed models of the QPOs themselves, the problem is overdetermined
so the underlying theories can be tested in critical ways.
\subsection{Equation of state of dense matter}
Nearly coherent oscillations at $\sim$ 300 Hz or $\sim$ 600 Hz have
been observed during type I X-ray bursts from about 10 NS so far (see
Strohmayer 1998 for a review).  These oscillations are probably
caused by rotational modulation of a hot spot on the stellar surface.
The emission from the hot spot is affected by gravitational light
deflection and Doppler shifts (e.g. Miller \& Lamb 1998).  With
XEUS, the oscillation will be detected within one cycle. The
composition and properties of the NS cores have been the subject of
considerable speculation, and remain a major issue in modern physics:
at the highest densities, matter could be composed of pion or kaon
condensates, hyperons, quark matter, or strange matter.  By fitting
the waveform, it will be possible to investigate the spacetime around
the NS, and simultaneously constrain its mass and radius, and hence
determine the equation of state of its high density core (see e.g.
Nath et al. 2002).

\subsection{Additional science}
A fast X-ray timing capability would allow XEUS to investigate the
physics of a wide range of astrophysical sources, such as accreting
millisecond pulsars, microquasars, X-ray pulsars, dippers, CVs,
novae, soft gamma-ray repeaters, anomalous X-ray pulsars.
For instance, there are only three accreting millisecond pulsars known so far; the first one discovered being the famous SAXJ1808-3658 (Wijnands \& van der Klis 1998; Chakrabarti \& Morgan 1998).  Its properties suggest that all NS systems should show pulsations at some
level.  In most models, pulse amplitudes cannot be suppressed below
$\sim0.1$\% (RMS) without conflicting with spectroscopic or QPO
evidence.  This is a factor of 10 above the sensitivity XEUS could achieve (millisecond pulsations at the 0.01\% RMS level would be detected in 1000 seconds in Sco X-1).  Detection of such pulsations in objects also showing kHz QPOs and burst
oscillations would immediately confirm or reject several models for
these phenomena involving the NS spin (e.g.
Miller et al. 1998).  In addition, it has been suggested
that such objects could be among the brightest gravitational radiation
sources in the sky, emitting a periodic gravitational wave signal at
the star's spin frequency (Bildsten 1998).  Undirected
searches in frequency space for such radiation lose sensitivity
because of statistical considerations.  Independently measuring the
spin period very accurately would therefore be of great importance for
periodicity searches with gravitational wave antennae (e.g.
Brady et al. 1997).

Another important area of astrophysics where the fast X-ray timing capability could contribute concerns microquasars. In these systems, the link between the very fast disk transitions
observed in X-rays and the acceleration process could be studied on
very short time scales, allowing the non steady state disk properties
and their link to the formation of relativistic jets to be explored
(Belloni et al. 1997; Fender et al. 1999).  This would
be of direct relevance to understanding the properties of AGNs, where
presumably similar jet formation mechanisms operate on a much larger
scales. In addition, through time-resolved spectroscopic observations,
the spacetime close to the black holes could be probed using the
variability of the iron K$\alpha$ line.

\section{XEUS sensitivity for timing studies}
\label{sensitivity} For the sensitivity computations, we have
assumed the energy response of the XEUS mirrors as given in the most recent
report of the telescope working group
(Aschenbach et al. 2001).  We have further assumed the
proposed high energy extension in which the inner mirror shells of the
telescope are coated with supermirrors (the effective area is thus
$\sim 20000$ cm$^2$ at $\sim 9$ keV and $\sim 1700$ cm$^2$ at 30 keV).
Finally we have assumed that the timing detector is made of 300 microns of Silicon lying above 2 mm of CdZnTe (see below).  Table \ref{dbarret-E1_tab:tab1} gives the count rates expected from some sources.

\begin{table}[h]  
\caption{Examples of total count rates above 0.5 keV and above 10
  keV (C$_{\rm E >10 keV}$) in kcts/s. The spectrum of Sco X-1 which is variable is such that it would produce 60 kcts/s in the RXTE/PCA (2.5--30 keV). The X-ray burst input spectrum is a blackbody of 1.5 keV with a normalization yielding an Eddington luminosity at 8.5 kpc. SAXJ1808-3659 is the millisecond pulsar taken at the peak of its 1996 outburst. }
  \label{dbarret-E1_tab:tab1}
  \begin{center}
    \leavevmode
    \footnotesize
    \begin{tabular}[h]{lccc}
      \hline \\ [-10pt]
      Source name & XEUS-1 & XEUS-2 & C$_{\rm E >10 keV}$ \\
      \hline \\ [-10pt]
      Crab & 250 & 800 & 5 \\
      Sco X-1 & 1200 & 3800 & 10 \\
      GC X-ray burst & 120 & 220 & 2 \\
     SAXJ1808-3659 & 30 & 130 & 0.3 \\
      \hline \\
      \end{tabular}
  \end{center} \end{table}

Let us now compute the sensitivity for QPO and coherent signal
detections. First the signal to noise ratio $n_\sigma$ at
which a QPO is detected in a photon counting experiment is approximately:
$$n_\sigma = {1\over2}{S^2\over
  B+S}r_S^2\left( T\over\Delta\nu \right)^{1/2}$$
where $S$ and $B$ are
source and background count rate, respectively, $r_S$ is the (RMS)
amplitude of the variability expressed as a fraction of $S$, $T$ the
integration time and $\Delta\nu$ the bandwidth of the variability. The
bandwidth $\Delta\nu$ is related to the coherence time $\tau$ of a QPO
as $\Delta\nu=1/\tau$. On the other hand, for a coherent signal
($T$$<$$1/\Delta\nu$), the more familiar exponential detection regime
applies, with false-alarm probability
$\sim$$\exp[-{S^2r_S^2T/2(B+S)}]$.

\begin{figure}[!t]
\centerline{\psfig{file=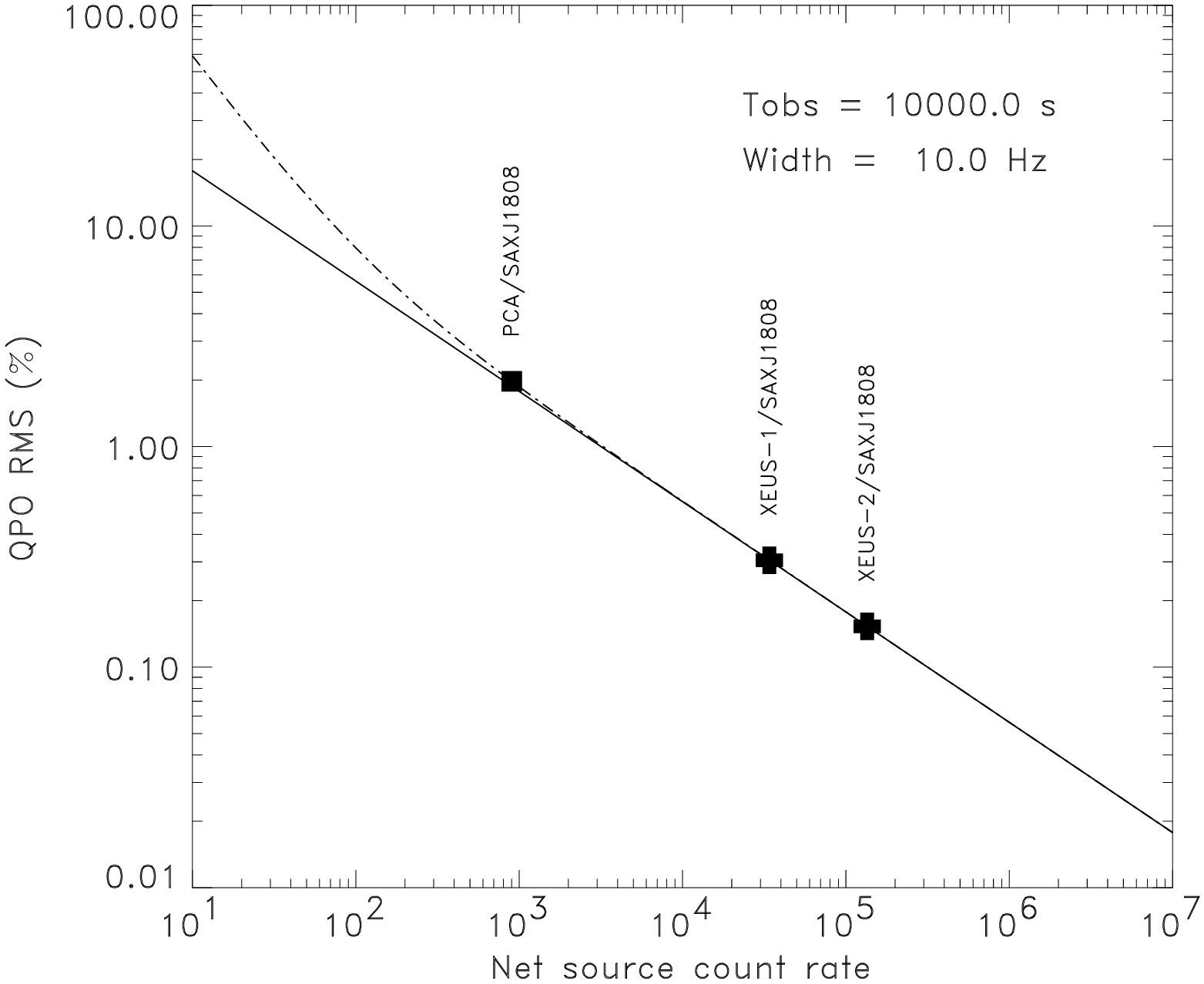,width=7.50cm}\psfig{file=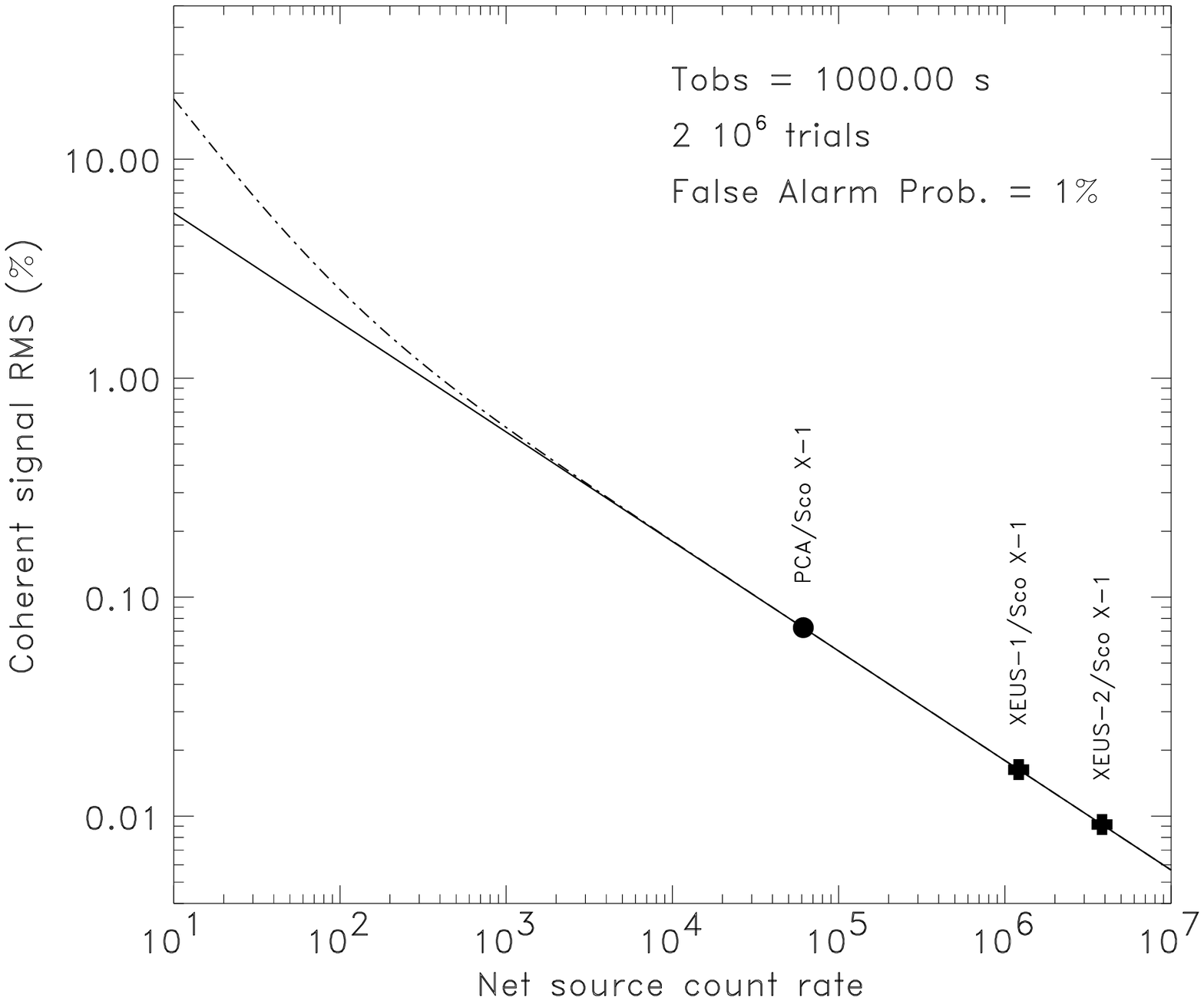,width=7.50cm}}
\caption{left: Comparison between the XEUS (solid line) and
  RXTE/PCA (dot-dashed line) sensitivity for QPO detection ($5\sigma$
  in 10 ksec, signal width 10 Hz). An illustrative example is provided
  by the millisecond pulsar for which RXTE failed to detect QPOs. As
  can be seen, a factor of $\sim 10$ improvement in sensitivity over
  the RXTE/PCA is obtained with XEUS. Right: Comparison between the XEUS (solid line) and
   RXTE/PCA (dot-dashed line) sensitivities for coherent signal detection
   (1 ksec). The detection level corresponds to a false alarm
   probability of 1\% for $2\times 10^6$ trials. So far, no pulsations
   have been detected from Sco X-1.  The XEUS-1 sensitivity is 10
   times better than the current RXTE/PCA sensitivity, and failure to detect pulsations at this level would demand major revision of our current ideas about low-mass X-ray binaries.}
\label{didier_barret_fig:fig1}
\end{figure}

From the above formulae, assuming B $\sim 0$ appropriate for XEUS, one
can estimate the RMS amplitude corresponding to a $5\sigma$ QPO
detection as a function of the source count rate (Figure
\ref{didier_barret_fig:fig1} left). Similarly one can compute the RMS for the
detection of a coherent signal at a given false alarm probability
(Figure \ref{didier_barret_fig:fig1} right). These two plots demonstrate that
with its huge collecting area XEUS provides an order of magnitude
sensitivity improvements in timing studies over RXTE. The sensitivity reached is such that QPOs could be detected within their coherence times and the oscillations detected within one cycle. The scaling of the above formula implies also that a QPO detected at $5\sigma$ with the PCA would be detected at $100\sigma$ with XEUS-1. Similarly, XEUS-1 will detect signals at the same level of significance as the PCA but for an observing time 100 times shorter.

\section{Detector implementation}
\subsection{Science requirements}
The detector needs to be able to handle up to 3 Mcts/s (XEUS-1) and 10
Mcts/s (XEUS-2) (equivalent to a 10 Crab source, see Table 1) with a
timing resolution of $\sim 10 \mu$s and a deadtime less than $\sim
1$\%. In addition, the detector energy range should match closely the high energy
response of the mirrors.

\subsection{Silicon Drift Detector}
In the current XEUS detector baseline, the Wide Field Imager (WFI) has
the highest count rate capabilities.  However, even in the most
optimistic case, it will only be able to provide timing information up
to 500 kcts/s (by using a fast window mode).  This means that an
alternative solution should be considered.  

\begin{figure}[!ht]
 \centerline{\psfig{file=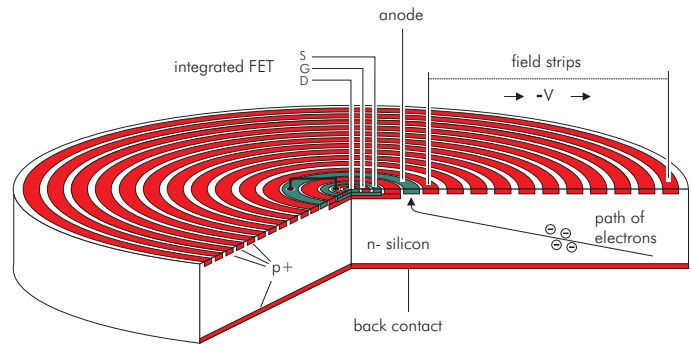,width=9.00cm}}
\caption{Schematic cross section of a cylindrical
    Silicon Drift Detector (SDD). Electrons are guided by an electric
    field towards the small collecting anode located at the center of
    the device. The first transistor of the amplifying electronics is
    integrated on the detector ship (drawing kindly provided by P.
    Lechner).}  \label{didier_barret_fig:fig2} \end{figure}

Among the fast X-ray
detectors currently available, Silicon Drift Detectors (SDDs) are the
most promising (Str{\" u}der 2000; Lechner et al. 2001).  The SDD is a completely depleted volume of silicon in which an arrangement of increasingly negative biased rings drive the electrons
generated by the impact of ionising radiation towards a small readout
node in the center of the device. The time needed for the electrons to
drift is much less than 1 $\mu$s.  The main advantage of SDDs over
conventional PIN diodes is the small physical size and consequently
the small capacitance of the anode, which translates to a capability
to handle high count rates simultaneously with good energy resolution.
To take full advantage of the small capacitance, the first transistor
of the amplifying electronics is integrated on the detector chip (see
Fig. \ref{didier_barret_fig:fig2}). The stray capacitance of the
interconnection between the detector and amplifier is thus minimized,
and furthermore the system becomes practically insensitive to
mechanical vibrations and electronic pickup.

\begin{figure}[!ht]
 \centerline{\psfig{file=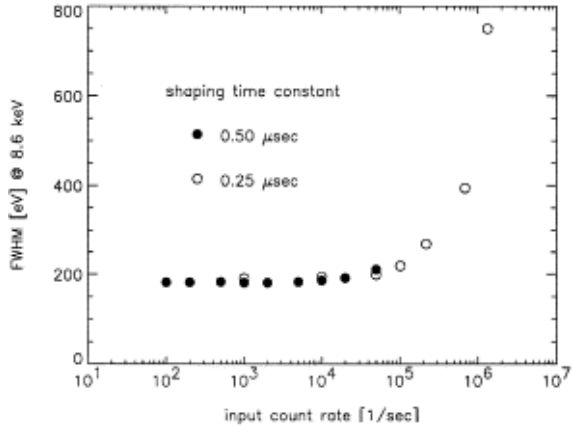,width=7.500cm}}
\caption{Dependence of energy resolution of a SDD with integrated FET on the rate of incoming photons ( Lechner et al. 2001).}  \label{didier_barret_fig:fig3} \end{figure}

Energy resolution of better than $\sim 200$ eV (at 6 keV,
equivalent to a low energy threshold $\sim 0.5$ keV) is readily
achieved with modest cooling (-20\degr C) for count rates below 10$^5$
cts/s (e.g. Lechner et al. 2001, see Fig. \ref{didier_barret_fig:fig3}). 

With such a low energy threshold, the fast timing capability would explore completely new windows of X-ray timing by getting below $\sim 2.5$ keV (current threshold of RXTE-like proportional counters). Such device would also allow the investigations of the frequency domain up to $\sim 10^4$ Hz, where signals have been predicted from accreting neutron stars (Sunyaev \& Revnivtsev 2000).  

\subsection{Deadtime and implications}
For timing studies, deadtime is always a critical issue. Deadtime will include contributions from the signal rise time, the charge sensitive amplifier, the shaping amplifier. The first two of these can be very short, and the limiting contribution is that of the amplifier, where a trade-off between speed and energy resolution is necessary. Shaping time constants as short as 50 nanoseconds (ns) have been found to be usable (Str{\" u}der 2000). This translates to a minimum feasible deadtime of $\sim 100$ ns. Using currently available devices and pipelining techniques, the analog-digital conversion stage is not a limiting factor at these speeds.

\begin{figure}[!h]
 \centerline{\psfig{file=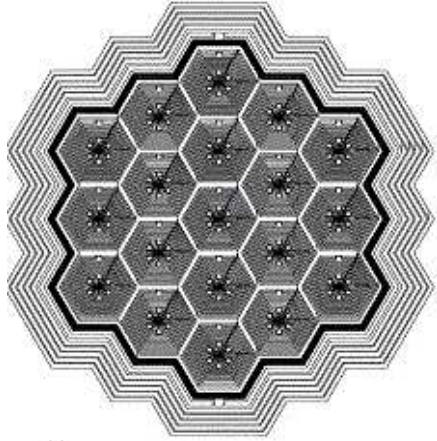,width=5.7500cm}}
\caption{SDD array made of 19 hexagon cells of 5 mm$^2$ (Lechner et al. 2001). The overall size of the detector is just about 1 cm$^2$.}  \label{didier_barret_fig:fig4} \end{figure}

A 100 ns deadtime per event corresponds to a 1\% deadtime for a source producing $10^5$ cps/s. To handle $10^6$ cps/s, one must therefore distribute the focal beam over  $\sim 10$ pixels. The best and easiest solution could be a detector made of an ensemble of about $\sim 10$ separate SDDs on a single wafer. Such SDD arrays already exist, as shown in Figure \ref{didier_barret_fig:fig4}. {\it This detector should therefore be operated out of focus}. For XEUS-1, the out of focus distance is of the order of 10 cm. This could be accomplished either by a mechanical construction, or by changing the distance between the detector and mirror spacecraft. Although this will require a careful study, both solutions appear to be feasible within the current XEUS mission design. The requirements in terms of real estate on the detector spacecraft are not constraining, in particular because no complicated cooling systems will be necessary. Finally, the SDD array could be easily implemented on the side of the wide field imager chip. 

\subsection{Radiation hardness}

The detector will be exposed to high radiation doses and one must therefore consider its radiation hardness. The main limitation in the maximum acceptable dose arises from the JFET connected to the collecting anode on the back of the device (Leutenegger et al. 2001). High energy photons absorbed in the transistor region increase the amount of oxyde charge and interface traps, thus reducing the charge carrier lifetimes, and thus contributing to increase the leakage current. Laboratory measurements indicate however that a 300 micron thick SDD survives a radiation dose of $\sim 10^{13}$ incoming high energy photons (E$>12$ keV) (Leutenegger et al. 2001). This is  equivalent to a continuous exposition of 3 years at 10$^5$ photons/s. Similarly, the detector will be exposed to particles at a moderate rate (1-2 particles/cm$^2$/s). Again, the XMM-Newton EPIC PN cameras which have similar detector technology have not shown any degradations in performance in space (Str\"uder et al. 2001). So, the device selected can be clearly considered as radiation hard.

\subsection{High energy extension}
As mentioned above, a high energy extension (above 40 keV) is proposed for the mirrors (Aschenbach et al. 2001). SDDs are currently produced with thicknesses of 300 microns, which is adequate to cover the energy range below 10 keV. Although there are on-going efforts to make thicker devices, the best match of the high energy response of the mirrors will require the SDD array to be associated with a higher density  detector located underneath. Among the potential high energy semi-conductor detectors, CdZnTe as the one presented in Budtz-J{\o}rgensen et al. (2001) stands today as a very promising solution. Such a detector would both ensure the overlap in energy range with the SDD array, and provide a flat energy response up to $\sim 80$ keV and 10 microsecond timing resolution (Budtz-J{\o}rgensen et al. 2001). 

\subsection{Telemetry and data handling}

The goal is to send to the ground the time and energy information of every photon. For most sources, data compression will make this possible (within a 2 Mbits/s data rate) without compromising either time or energy resolution. For the very brightest sources, this can still be done with a restricted number of energy channels.

\section{Conclusions} 
Probing strong gravity fields, constraining the equation of state of dense matter, and more generally studying the brightest X-ray sources of the sky with fast X-ray timing could be achieved with XEUS with the addition of a fast X-ray timing capability in the focal plane. This would greatly extend the capabilities of XEUS at low cost, with little impact on the primary objectives of the mission, which remain the spectroscopy of the most distant X-ray sources of the Universe. Current detector technology meet the science requirements for the timing of the brightest X-ray sources. The most attractive detector implementation would consist of an array of small size silicon drift detectors operated out of focus.
\begin{acknowledgements}
We are grateful to the following colleagues
  who are supporting the proposal for a fast X-ray timing capability
  in XEUS: J.L. Atteia, T. Belloni, H. Bradt, L. Burderi, S. Campana,
  A. Castro-Tirado, D. Chakrabarty, P. Charles, S. Collin, S. Corbel,
  C. Done, G. Dubus, M. Gierlinski, J. Grindlay, A. Fabian, R. Fender,
  E. Gourgoulhon, J.M. Hameury, C. Hellier, W. Kluzniak, E. Kuulkers,
  S. Larsson, J.P. Lasota, T. Maccarone, D. de Martino, K. Menou, C. Miller, F.
  Mirabel, M. Nowak, J.F. Olive, S. Paltani, R. Remillard, J. Rodriguez, R.
  Rothschild, T. di Salvo, R. Sunyaev, M. Tagger, M. Tavani, L. Titarchuk, G.
  Vedrenne, N.  White, R. Wijnands, J. Wilms, A. Zdziarski, W. Zhang.
  
  The authors are also grateful to A. Parmar, G. Hasinger and M. Turner for helpful information about the XEUS mission in general. 

 \end{acknowledgements}

\end{document}